\documentclass[conference]{IEEEtran}
\IEEEoverridecommandlockouts
\usepackage{amsmath, float, subcaption, graphicx, multirow, diagbox}

\graphicspath{ {figs/} }

\begin{document}
\title{A Machine Learning Approach to Optimal Inverse \\ 
Discrete Cosine Transform (IDCT) Design}
%\author{Anonymous IEEE ISCAS 2021 Submission}

\author{\IEEEauthorblockN{
Yifan Wang\IEEEauthorrefmark{1},
Zhanxuan Mei\IEEEauthorrefmark{1}, 
Chia-Yang Tsai\IEEEauthorrefmark{2},
Ioannis Katsavounidis\IEEEauthorrefmark{2} and 
C.-C. Jay Kuo\IEEEauthorrefmark{1}}
\IEEEauthorblockA{\IEEEauthorrefmark{1}University of Southern California, Los Angeles, California, USA}
\IEEEauthorblockA{\IEEEauthorrefmark{2} Facebook, Inc., Menlo Park, California, USA}}

\maketitle

\begin{abstract}

The design of the optimal inverse discrete cosine transform (IDCT) to
compensate the quantization error is proposed for effective lossy image
compression in this work. The forward and inverse DCTs are designed in
pair in current image/video coding standards without taking the
quantization effect into account.  Yet, the distribution of quantized
DCT coefficients deviate from that of original DCT coefficients. This
is particularly obvious when the quality factor of JPEG compressed
images is small. To address this problem, we first use a set of
training images to learn the compound effect of forward DCT,
quantization and dequantization in cascade. Then, a new IDCT kernel is
learned to reverse the effect of such a pipeline.  Experiments are
conducted to demonstrate that the advantage of the new method, which has
a gain of 0.11-0.30dB over the standard JPEG over a wide range of
quality factors. 

\end{abstract}

\IEEEpeerreviewmaketitle

\section{Introduction}\label{sec:introduction}

Many image and video compression standards have been developed in the
last thirty years. Examples include JPEG \cite{wallace1992jpeg},
JPEG2000 \cite{rabbani2002jpeg2000}, and BPG \cite{bpg} for image
compression and MPEG-1 \cite{brandenburg1994iso}, MPEG-2
\cite{haskell1996digital}, MPEG-4 \cite{pereira2002mpeg}, H.264/AVC
\cite{wiegand2003h264overview} and HEVC \cite{sullivan2012hevc} for
video compression. Yet, we see few machine learning techniques adopted
by them. As machine learning becomes more popular in multimedia
computing and content understanding, it is interesting to see how
machine learning can be introduced to boost the performance of
image/video coding performance. In this work, we investigate how machine
learning can be used in the optimal design of the inverse discrete
cosine transform (IDCT) targeting at quantization error compensation. 

Block transforms are widely used in image and video coding standards for
energy compaction in the spatial domain.  Only a few leading DCT
coefficients have larger magnitudes after the transformation. Furthermore, a
large number of quantized coefficients become zeros after quantization.
These two factors contribute to file size reduction greatly.  The 8x8
block DCT \cite{ahmed1974dct} is used in JPEG. The integer cosine
transform (ICT) is adopted by H.264/AVC for lower computational
complexity. All forward and inverse kernels are fixed in these
standards. 

Research on IDCT has primarily focused on complexity reduction via new
computational algorithms and efficient software/hardware implementation.
For example, an adaptive algorithm was proposed in
\cite{pao1999modeling} to reduce the IDCT complexity. A
zero-coefficient-aware butterfly IDCT algorithm was investigated in
\cite{park2016zero} for faster decoding. As to effective
implementations, a fast multiplier implementation was studied in
\cite{atitallah2006optimization}, where the modified Loeffler algorithm
was implemented on the FPGA to optimize the speed and the area.  There
is however little work on analyzing the difference of DCT coefficients
before and after quantization.  In image/video coding, DCT coefficients
go through quantization and de-quantization before their inverse
transform. When the coding bit rate is high,
quantized-and-then-dequantized DCT coefficients are very close to the input
DCT coefficients. Consequently, IDCT can reconstruct input image patches
reasonably well. 

However, when the coding bit rate is low, this condition does not hold
any longer. The quantization effect is not negligible. The distribution
of quantized DCT coefficients deviates from that of
original DCT coefficients. The traditional IDCT kernel, which is derived
from the forward DCT kernel, is not optimal. In this work, we attempt to
find the optimal IDCT kernel. This new kernel is derived from the
training set using machine learning approach. Simply speaking, it
transforms quantized DCT coefficients back to spatial-domain pixel
values, which is exactly what IDCT is supposed to do. Our solution can
account for the quantization effect through the training of real world
data so that the quantization error can be reduced in decoding.  It can
improve the evaluation score and visual quality of reconstructed images
without any extra cost in decoding stage. 

The rest of this paper is organized as follows. The impact of
quantization on the IDCT is analyzed in Sec. \ref{sec:impact}. The new
IDCT method is proposed in Sec. \ref{sec:method}. Experiment results are
presented in Sec. \ref{sec:experiment}. Finally, concluding remarks and
future research directions are given in Sec.  \ref{sec:conclusion}. 

\section{Impact of Quantization on Inverse DCT}\label{sec:impact}

The forward block DCT can be written as
\begin{equation}\label{eq:DCT}
{\bf X} = K {\bf x},
\end{equation}
where ${\bf x} \in R^N$, $N=n\times n$, is an N-dimensional
vector that represents an input image block of $n \times n$ pixels, $K
\in R^{N \times N}$ denotes the DCT transform matrix, and ${\bf X} \in
R^N$ is the vector of transformed DCT coefficients. Rows of $K$ are
mutually orthogonal. Furthermore, they are normalized in principle.  In
the implementation of H.264/AVC and HEVC, rows of $K$ may not be
normalized to avoid floating point computation, which could vary from
one machine to the other.  Mathematically, we can still view $K$ as a
normalized kernel for simplicity.  Then, the corresponding inverse DCT
can be represented by
\begin{equation}\label{eq:IDCT}
{\bf x}_q = K^{-1} {\bf X}_q = K^T {\bf X}_q, 
\end{equation}
where $K^{-1}=K^T$ since $K$ is an orthogonal transform and ${\bf X}_q$
is the vector of quantized DCT coefficients. Mathematically, we have
\begin{equation}\label{eq:QDCT} 
{\bf X}_q= \mbox{DeQuan}(\mbox{Quan}({\bf X})),
\end{equation}
where $Quan(\cdot)$ represents the quantization operation and
$DeQuan(\cdot)$ represents the de-quantization operation.  If there
is little information loss in the quantization process (i.e., nearly
lossless compression), we have ${\bf X}_q \approx {\bf X}$.
Consequently, the reconstructed image block ${\bf x}_q$ will be close to
the input image block, ${\bf x}$.  Sometimes,  a smaller
compressed file size at the expense of lower quality is used. Then, the
difference between ${\bf X}_q$ and ${\bf X}$ can be expressed as
\begin{equation}\label{eq:error} 
{\bf e}_q = {\bf X} - {\bf X}_q.
\end{equation}

In decoding, we have ${\bf X}_q$ rather than ${\bf X}$. If we apply
the traditional IDCT (i.e., $K^{-1}=K^T$) to ${\bf X}_q$, we can 
derive the image block error based on Eqs. (\ref{eq:IDCT}), 
(\ref{eq:IDCT}) and (\ref{eq:error}) as
\begin{eqnarray} \label{eq:error_3}
{\bf e}_q = {\bf x} - K^{-1} \mbox{DeQuan}[\mbox{Quan}(K {\bf x})].
\end{eqnarray}
To minimize ${\bf e}_q$, we demand taht
\begin{equation}\label{eq:qIDCT} 
K_q^{-1} \approx \mbox{DeQuan} \odot \mbox{Quan} \odot K
\end{equation}
where $\odot$ denotes an element-wise operation. Since it is difficult
to find a mathematical model for the right-hand-side of Eq.
(\ref{eq:qIDCT}), we adopt a machine learning approach to learn the
relationship between quantized DCT coefficients, ${\bf X}_q$, and pixels
of the original input block, ${\bf x}$, from a large number of training
samples. Then, given ${\bf X}_q$, we can predict target ${\bf x}$. 

\section{Proposed IDCT Method}\label{sec:method}

We use image block samples of size $8\times 8$ to learn the optimal IDCT
matrix. The training procedure is stated below. 
\begin{enumerate}
\item Perform the forward DCT on each image block.
\item Quantize DCT coefficients with respect to quality factor $QF$.
\item Conduct linear regression between de-quantized DCT coefficients,
which are the input, and image pixels of input blocks, which are the
output, to determine optimal IDCT matrix $\hat{K}$. 
\end{enumerate}
Then, we will use $\hat{K}$, instead of $K^{-1}$ in Eq.
(\ref{eq:error_3}), for the IDCT task. Note that it is costly to train
$\hat{K}$ at every QF value.

For Step 2, it is important to emphasize that there is no need to train
and store $\hat{K}$ for every QF value, which is too costly. 
Instead, we can train and store several $\hat{K}$ for a small set of selected
quality factors. Then, according to the QF that is used to encode images, we can choose the K with the closest QF to decode them. This point will be elaborated in Sec.
\ref{sec:experiment}.

For Step 3, we first reshape the 2D layout of
image pixels and dequantized DCT coefficients (both of dimension $8
\times 8$) into 1D vectors of dimension $64$. Notation ${\bf x}_i$
denotes the flattened pixel vector of the $i$th image block and
\begin{equation}
{\bf X}_{q,i}=\mbox{DeQuan}[\mbox{Quan}(K {\bf x}_i)]
\end{equation}
denotes the flattened dequantized DCT coefficients of the $i$th image
block. Next, we form data matrix $P \in R^{64\times N}$ of image pixels
and data matrix $D \in R^{64\times N}$ of dequantized DCT coefficients.
Columns of matrix $P$ are formed by ${\bf x}_i$ while columns of matrix
$D$ are formed by ${\bf X}_{q,i}$, $i=1, \cdots, N$. Then, we can set up
the regression problem by minimizing the following objective function
\begin{equation}
\xi (\hat{K}) = \sum_{i=1}^N || {\bf e}_i ||^2,
\end{equation}
where $|| \cdot ||$ is the Euclidean norm and ${\bf e}_i$ is the column 
vector of error matrix
\begin{equation}
E=P - \hat{K} D,
\end{equation}
and where learned kernel, $\hat{K}$, is the regression matrix of
dimension $64 \times 64$.  Typically, the sample number, $N$, is
significantly larger than $64^2=4,096$. 

%%%%%%%%%%%%%%%%%%%%%%%%%%%%%%%%%%%%%%%%%%%%%%%%%%
\begin{figure}[tp]
    \centering
     \begin{subfigure}[b]{0.5\textwidth}
         \hbox{\hspace{1.6em}
         \includegraphics[width=0.8\textwidth]{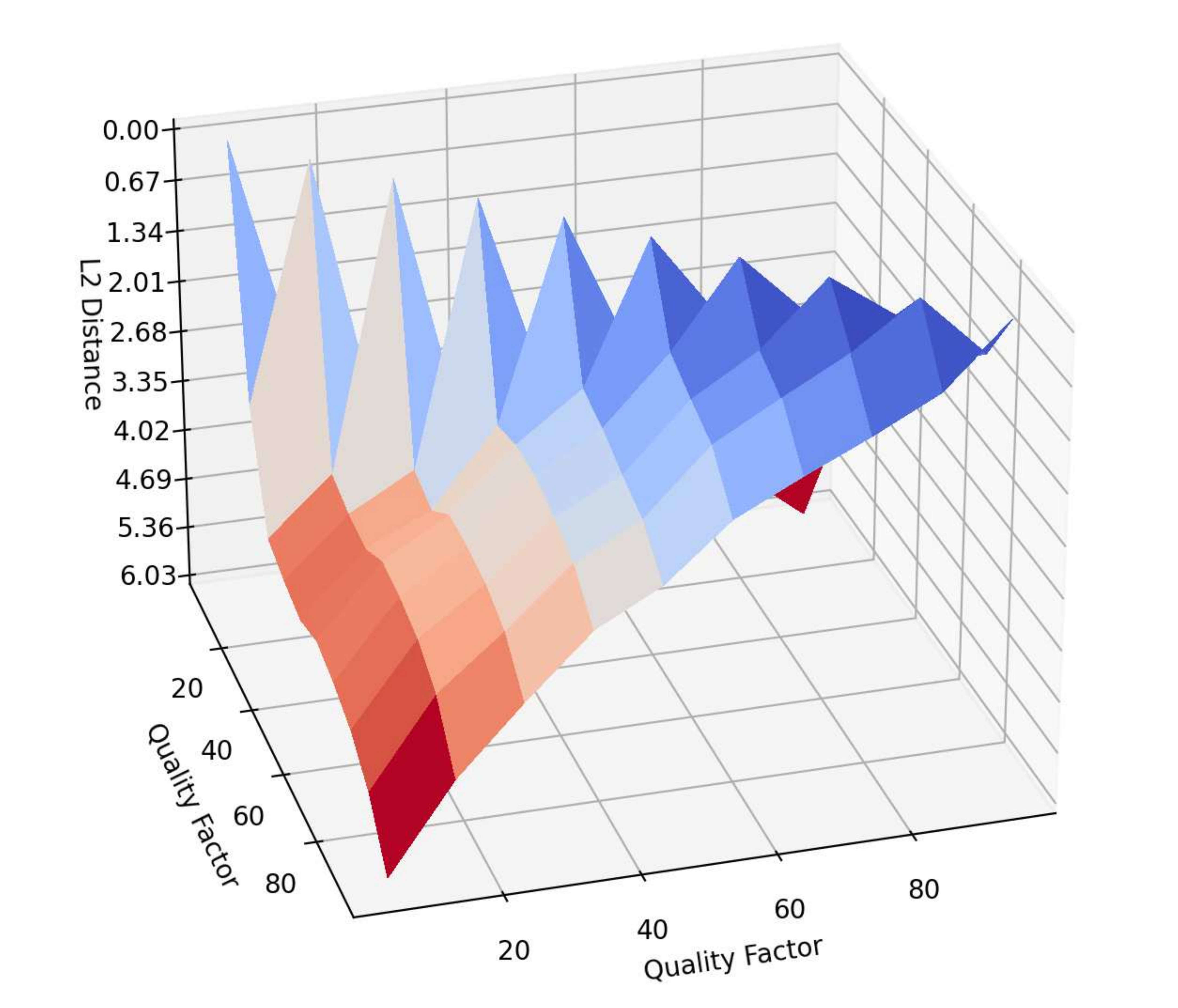}}
     \end{subfigure}
\caption{The $L_2$ norm of differences between kernels learned with 
different QFs.}\label{fig:l2}
\end{figure}
%%%%%%%%%%%%%%%%%%%%%%%%%%%%%%%%%%%%%%%%%%%%%%%%%%

%Jay: You may try a different viewing angle on this 3D plot. It does not look good as is.

%%%%%%%%%%%%%%%%%%%%%%%%%%%%%%%%%%%%%%%%%%%%%%%%%%
\begin{figure*}[htp]
     \centering
     \begin{subfigure}[b]{0.3\textwidth}
        \centering
         \includegraphics[width=0.99\textwidth]{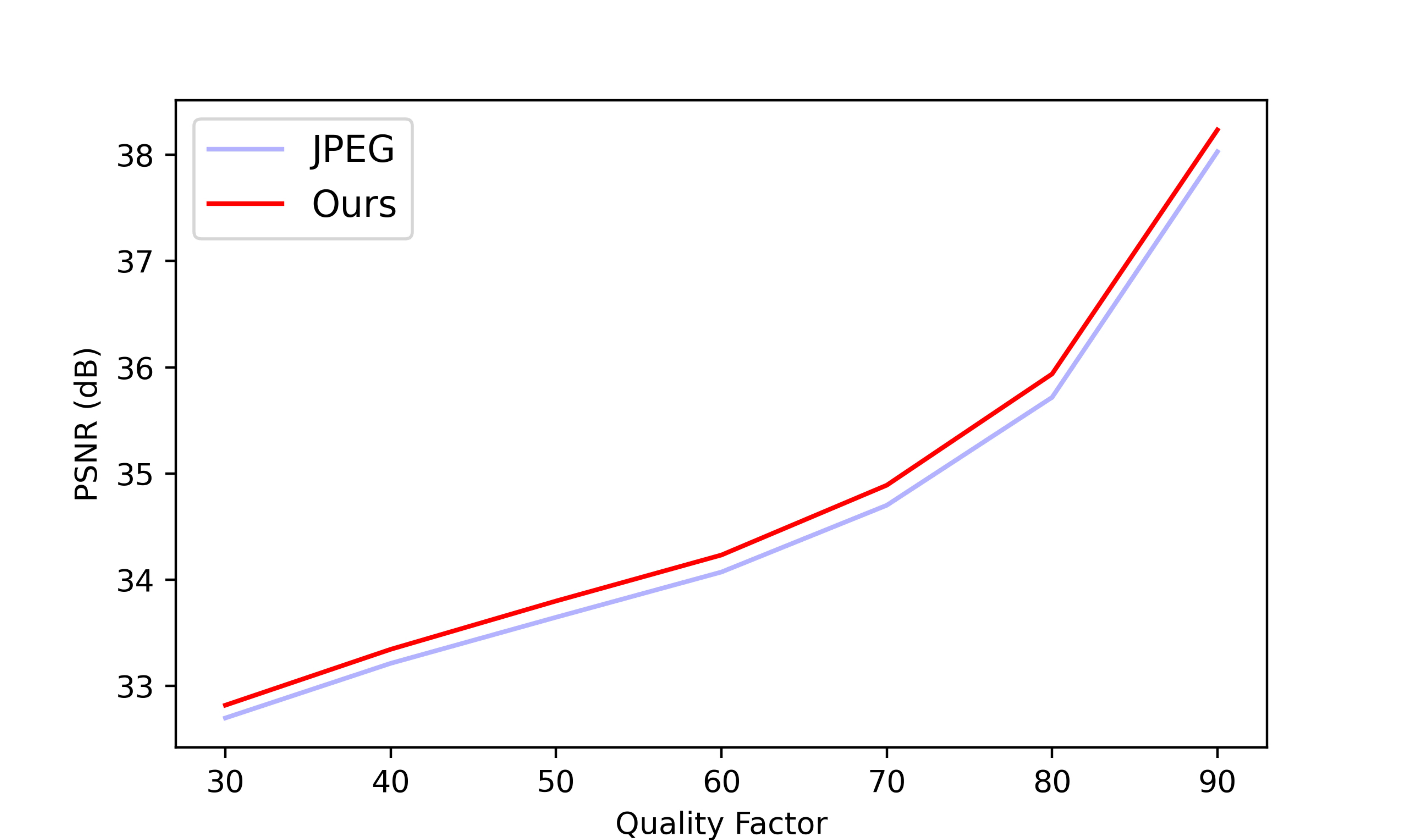}
         \caption{Kodak PSNR versus QF}
     \end{subfigure}
      \begin{subfigure}[b]{0.3\textwidth}
        \centering
         \includegraphics[width=0.99\textwidth]{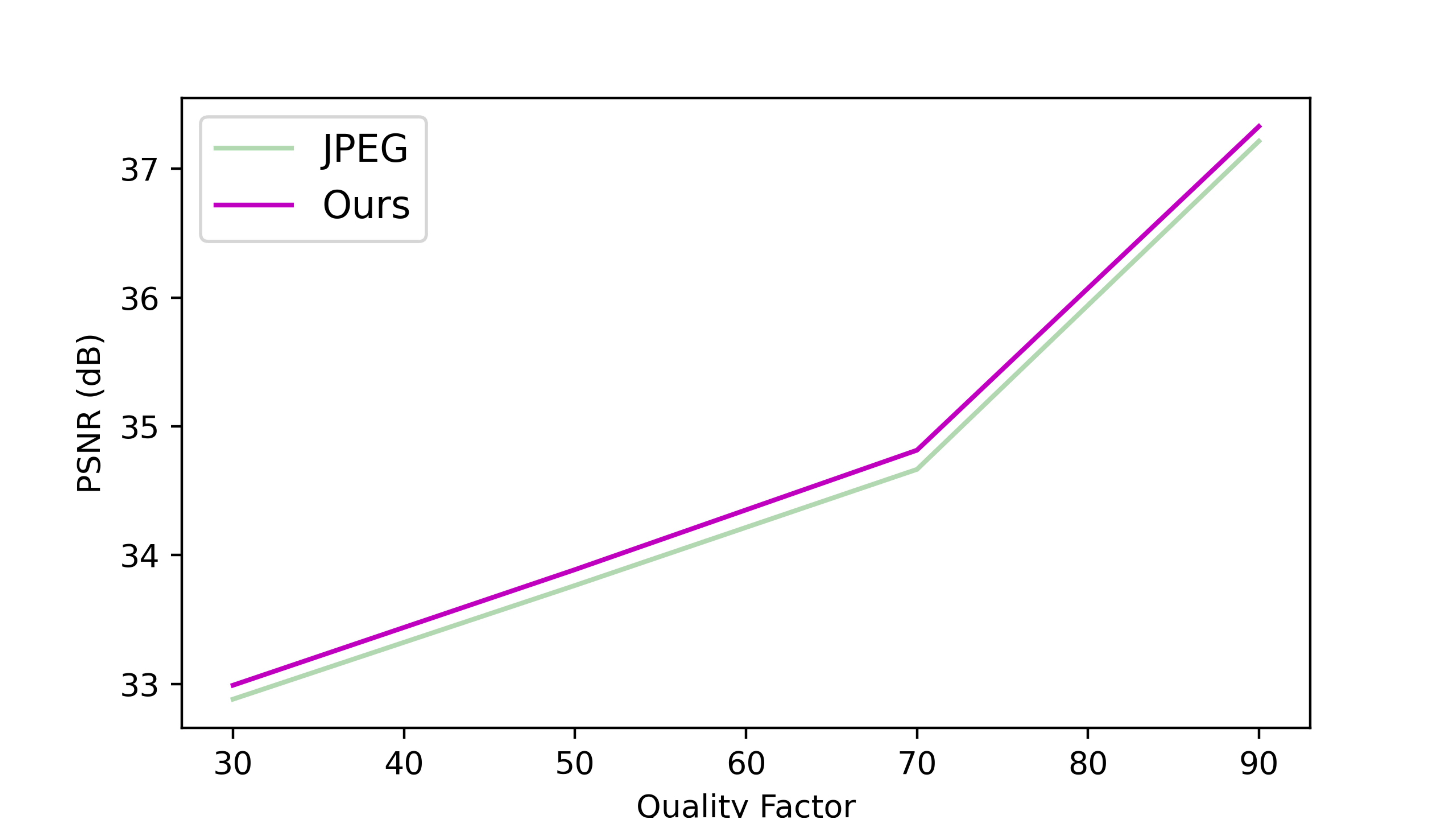}
         \caption{DIV2K PSNR versus QF}
     \end{subfigure}
        \begin{subfigure}[b]{0.3\textwidth}
        \centering
         \includegraphics[width=0.99\textwidth]{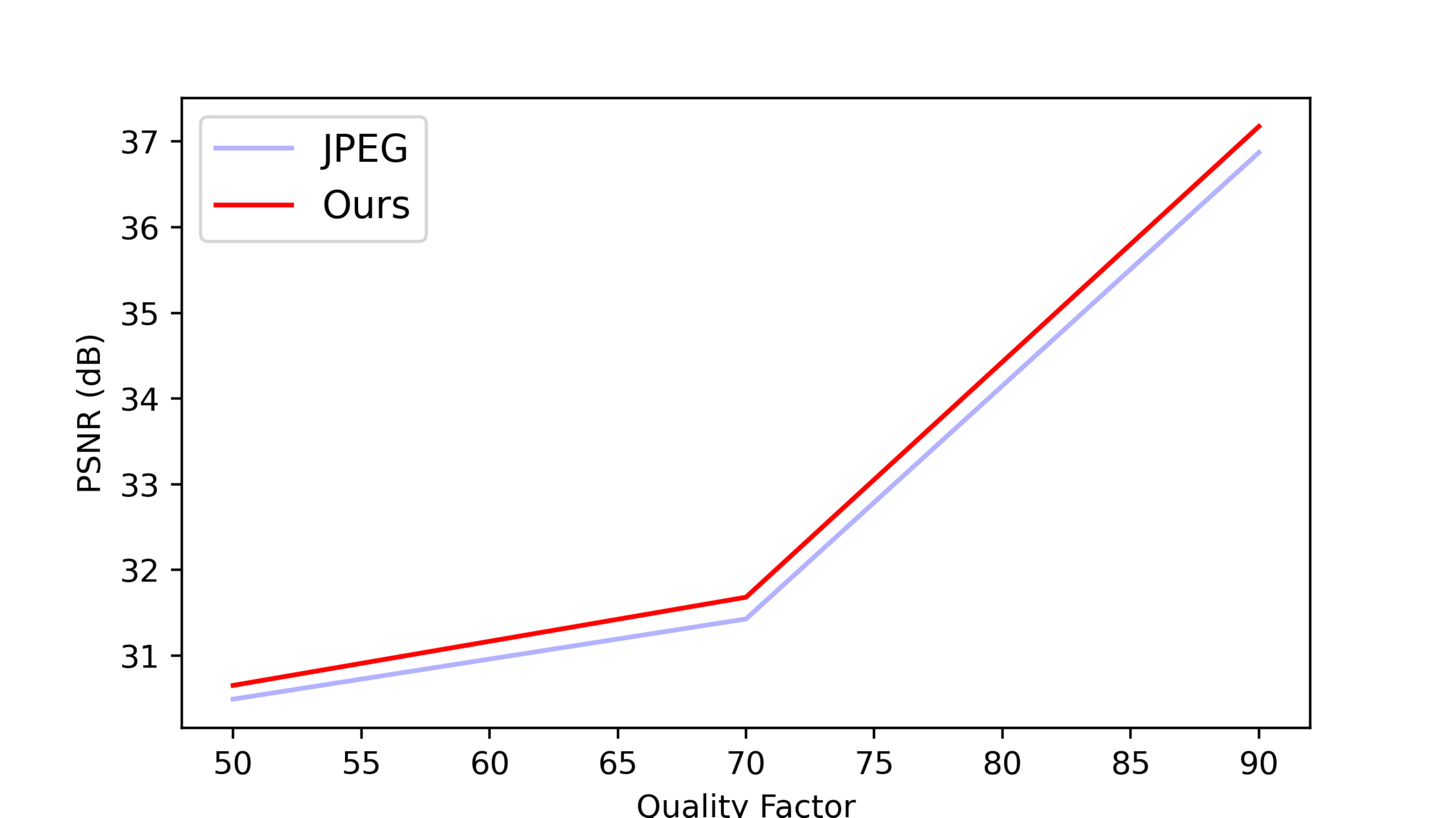}
         \caption{MBT PSNR versus QF}
     \end{subfigure}
     \begin{subfigure}[b]{0.3\textwidth}
        \centering
         \includegraphics[width=0.99\textwidth]{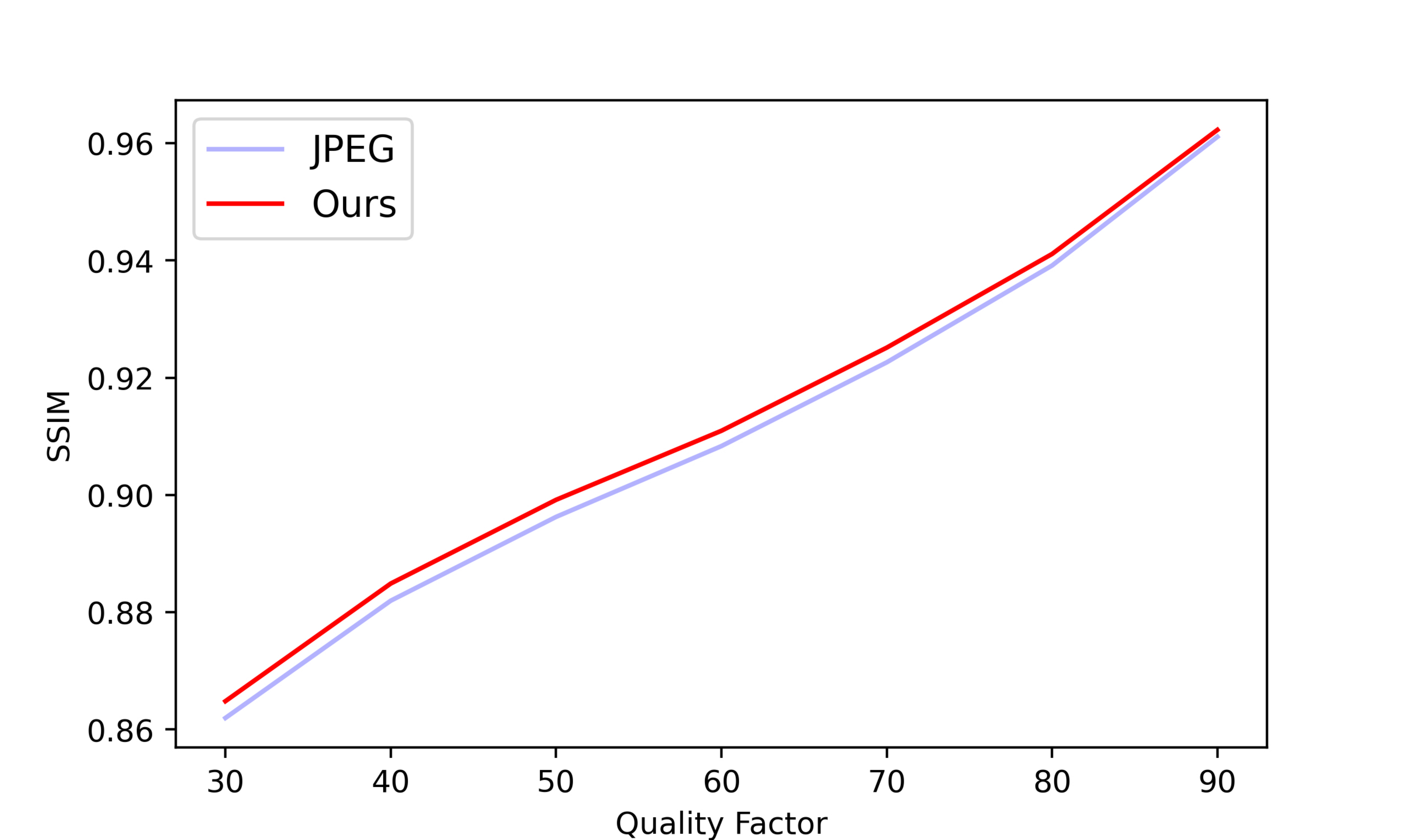}
         \caption{Kodak SSIM versus QF}
     \end{subfigure}
     \begin{subfigure}[b]{0.3\textwidth}
        \centering
         \includegraphics[width=0.99\textwidth]{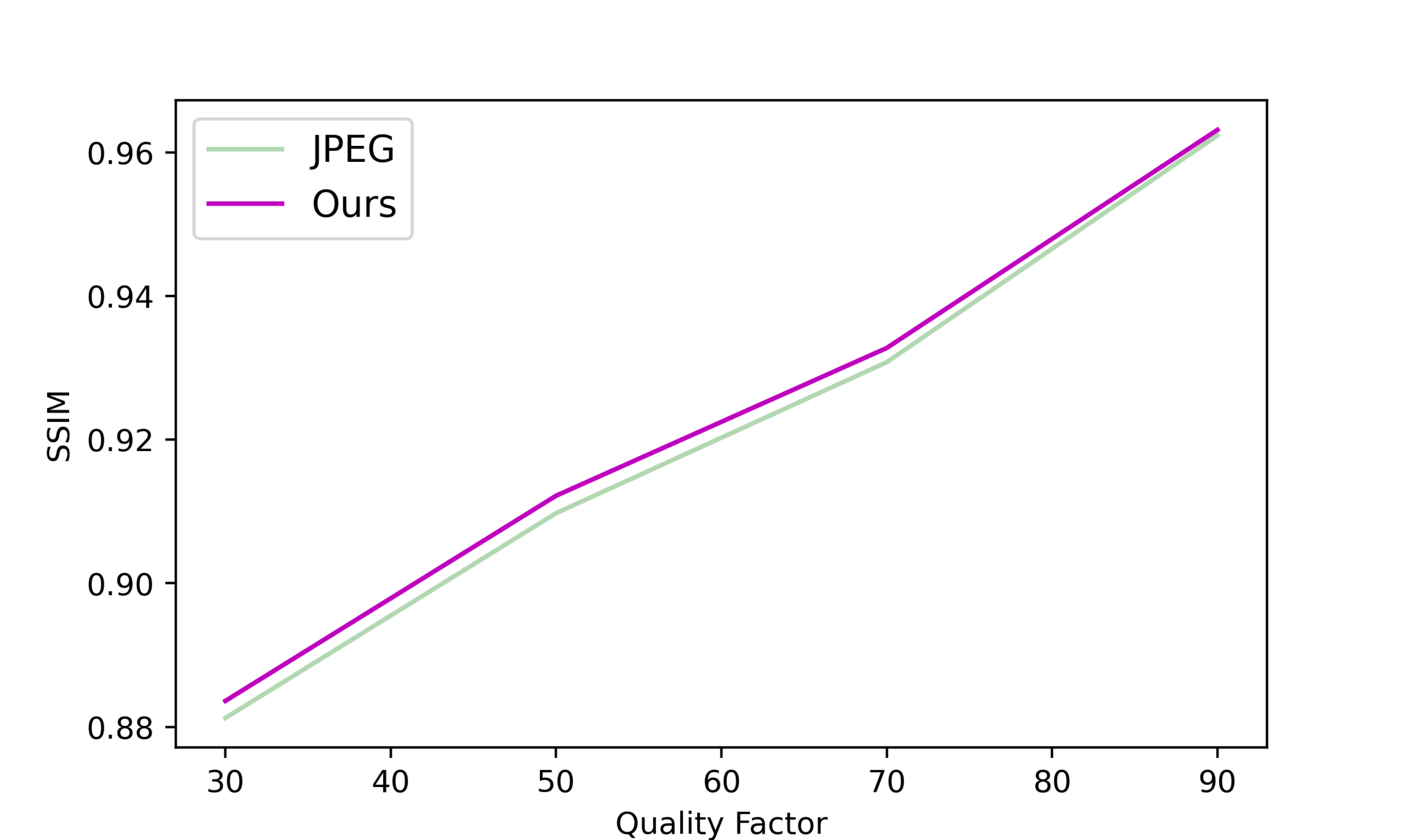}
         \caption{DIV2K SSIM versus QF}
     \end{subfigure}
     \begin{subfigure}[b]{0.3\textwidth}
        \centering
         \includegraphics[width=0.99\textwidth]{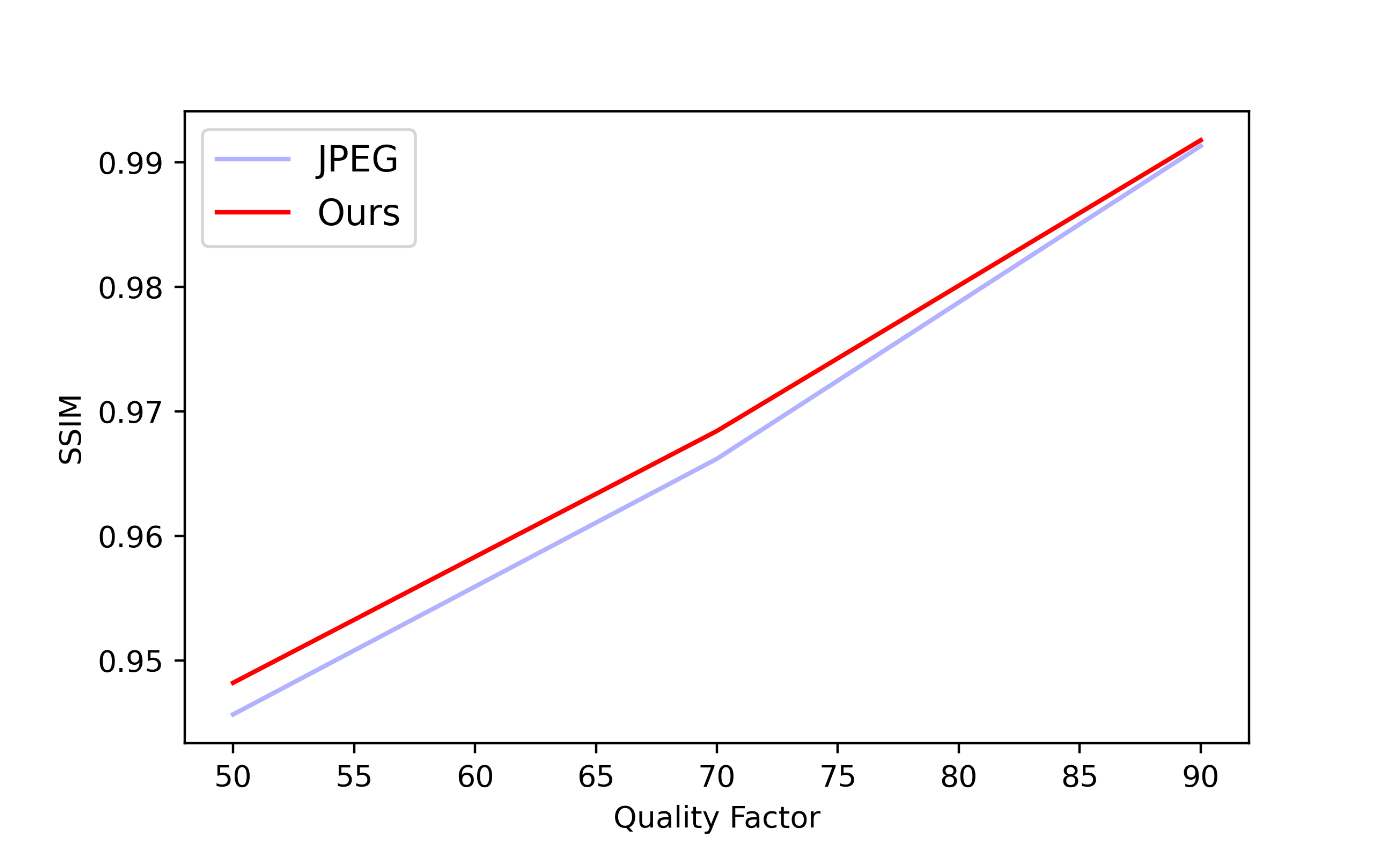}
         \caption{MBT SSIM versus QF}
     \end{subfigure}
\caption{Comparison of quality of decoded test images using the standard 
IDCT in JPEG and the proposed optimal IDCT for the Kodak, DIV2K and MBT 
datasets.}\label{fig:evaluation}
\end{figure*}
%%%%%%%%%%%%%%%%%%%%%%%%%%%%%%%%%%%%%%%%%%%%%%%%%%

%%%%%%%%%%%%%%%%%%%%%%%%%%%%%%%%%%%%%%%%%%%%%%%%%%%%
\begin{figure}[tp]
     \centering
     \begin{subfigure}[b]{0.5\textwidth}
        \centering
         \includegraphics[width=0.6\textwidth]{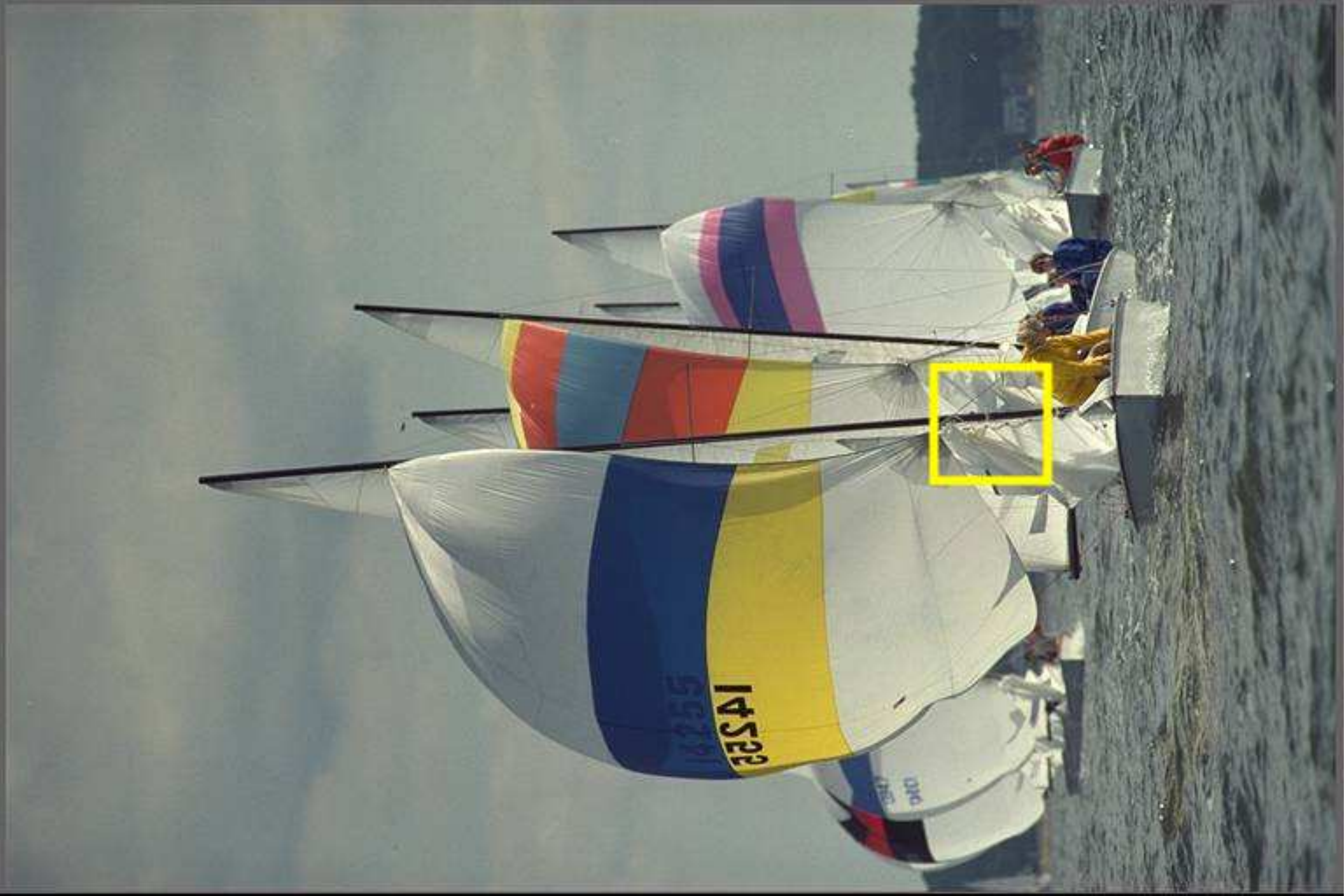}
         \caption{Input image from Kodak}
         \label{fig:kodak}
     \end{subfigure}
     \begin{subfigure}[b]{0.15\textwidth}
        \centering
         \includegraphics[width=0.8\textwidth]{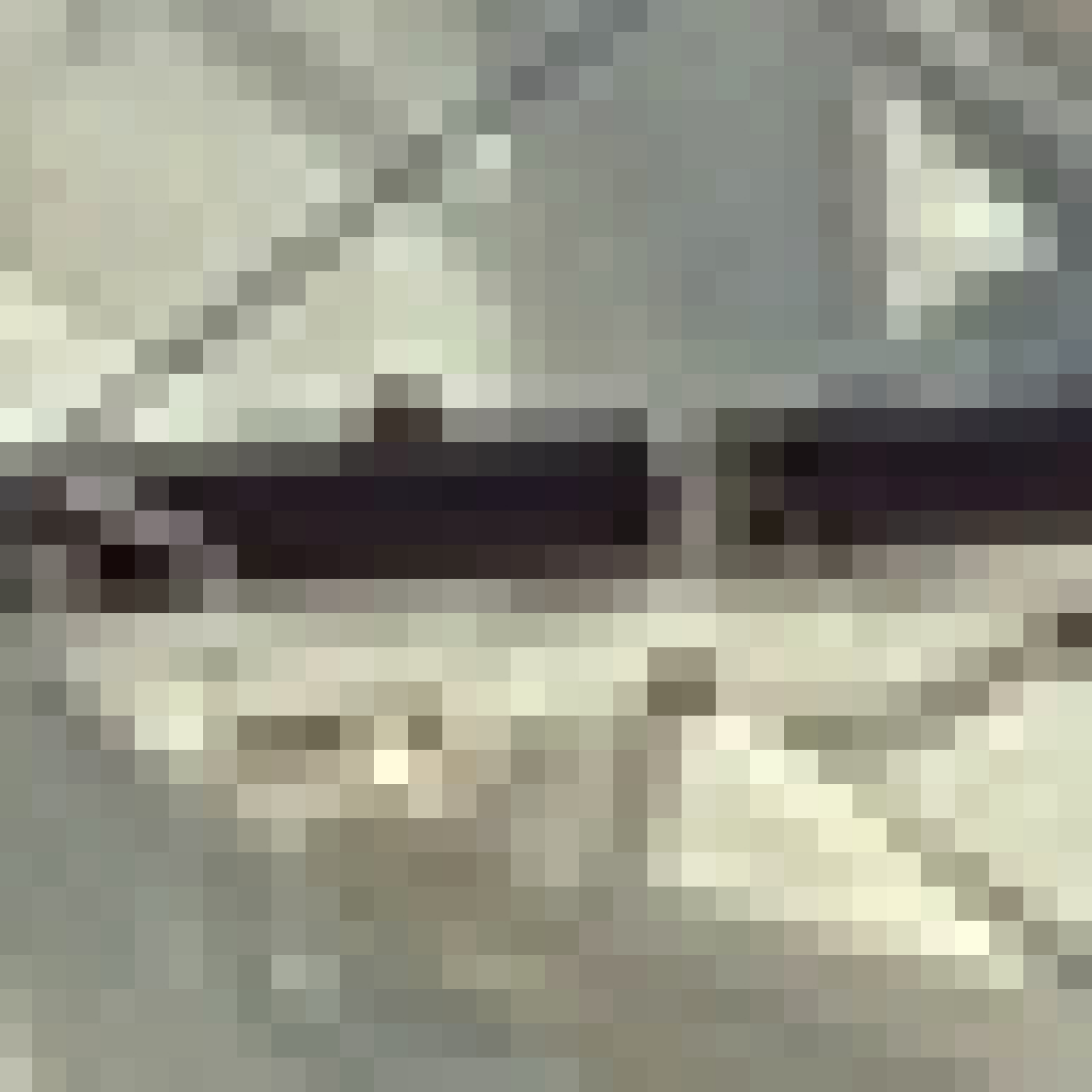}
         \caption{Zoom-in}
         \label{fig:kodakr}
     \end{subfigure}
     \begin{subfigure}[b]{0.15\textwidth}
        \centering
         \includegraphics[width=0.8\textwidth]{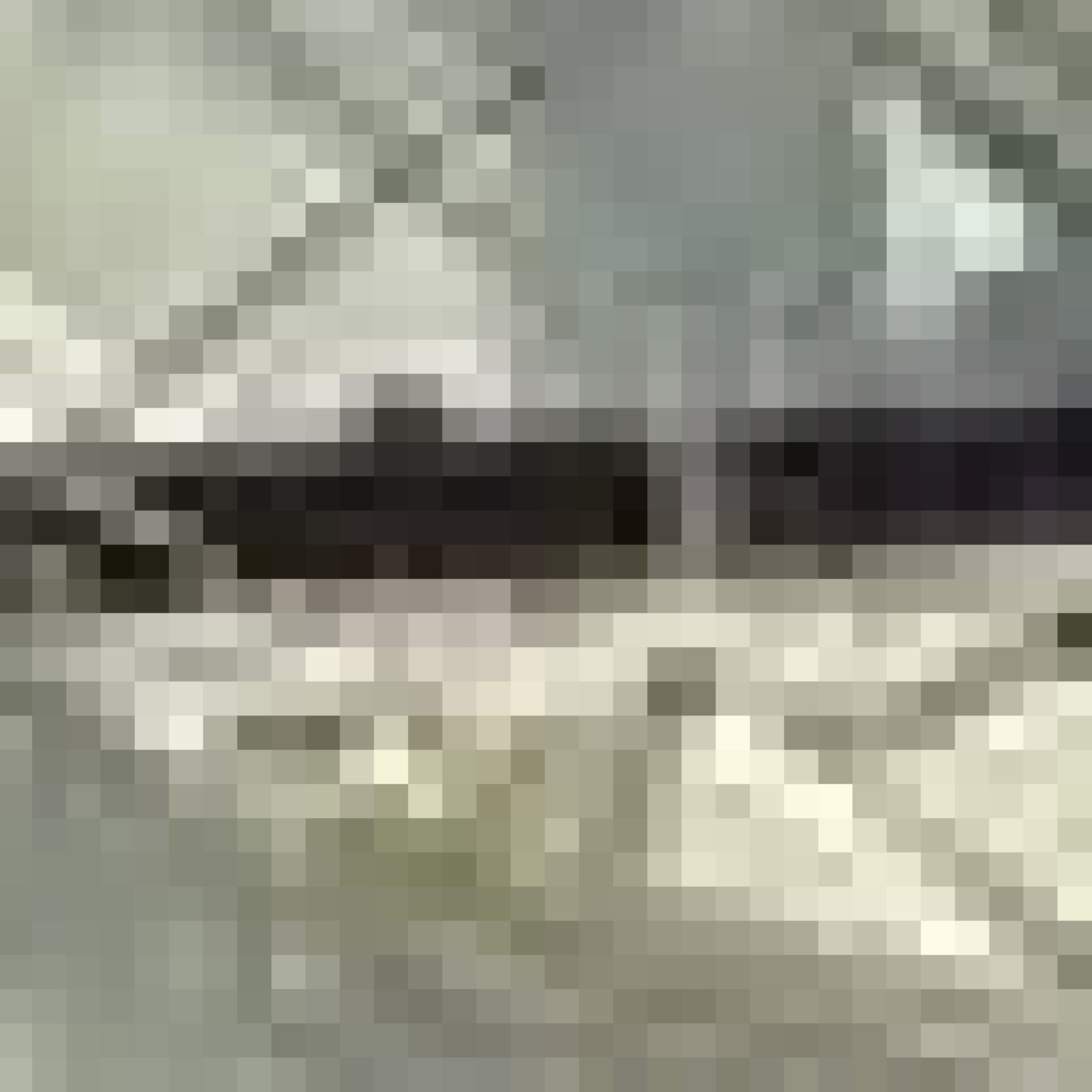}
         \caption{JPEG}
         \label{fig:kodakj}
     \end{subfigure}
     \begin{subfigure}[b]{0.15\textwidth}
        \centering
         \includegraphics[width=0.8\textwidth]{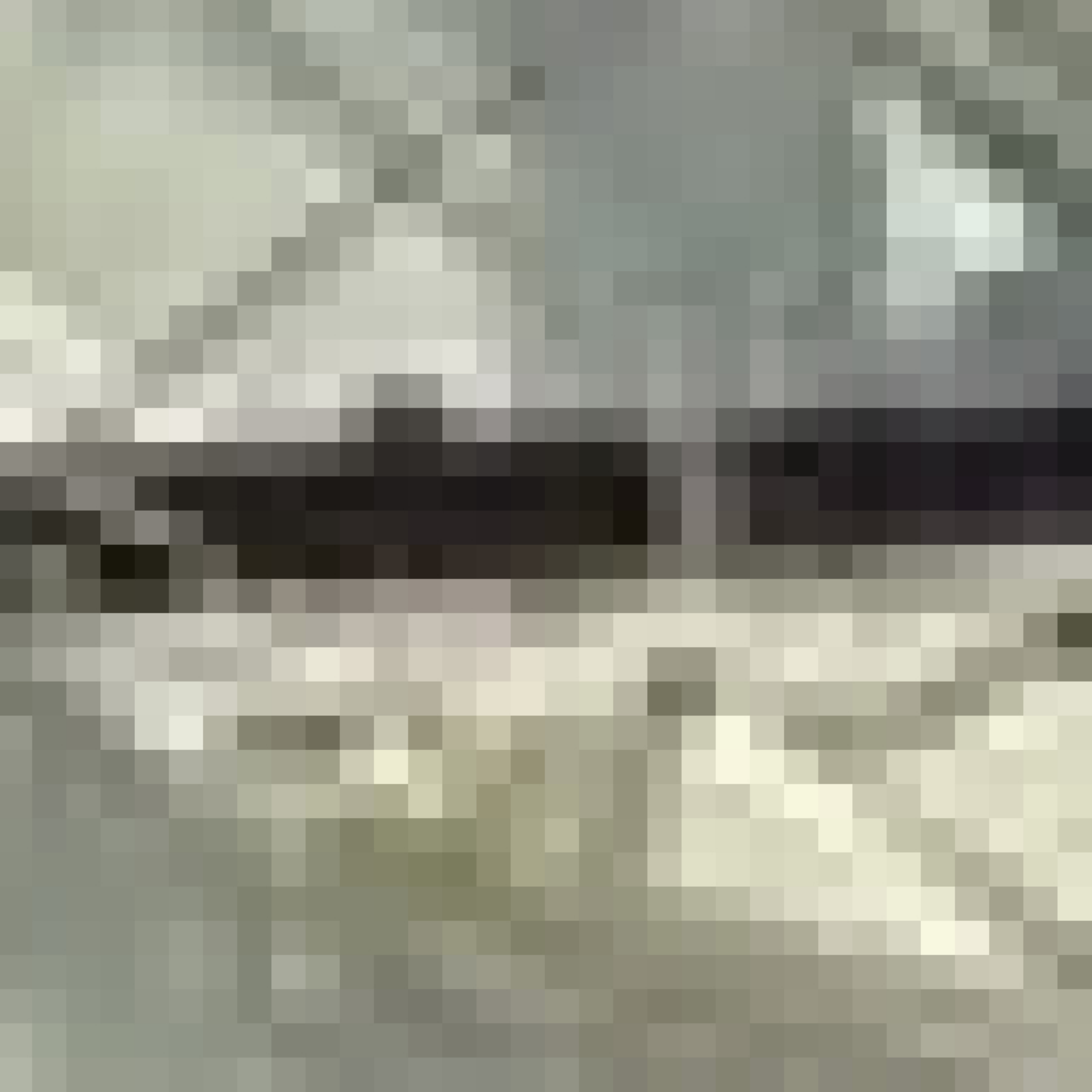}
         \caption{Ours}
         \label{fig:kodako}
     \end{subfigure}
\caption{Visualization of a zoom-in region of an input Kodak image, the
decoded region by JPEG and the proposed method with QF=70.}\label{fig:example_kodak}
\end{figure}
%%%%%%%%%%%%%%%%%%%%%%%%%%%%%%%%%%%%%%%%%%%%%%%%%%%%

%%%%%%%%%%%%%%%%%%%%%%%%%%%%%%%%%%%%%%%%%%%%%%%%%%
\begin{figure}[htp]
     \centering     
    \begin{subfigure}[b]{0.5\textwidth}
        \centering
         \includegraphics[width=0.75\textwidth]{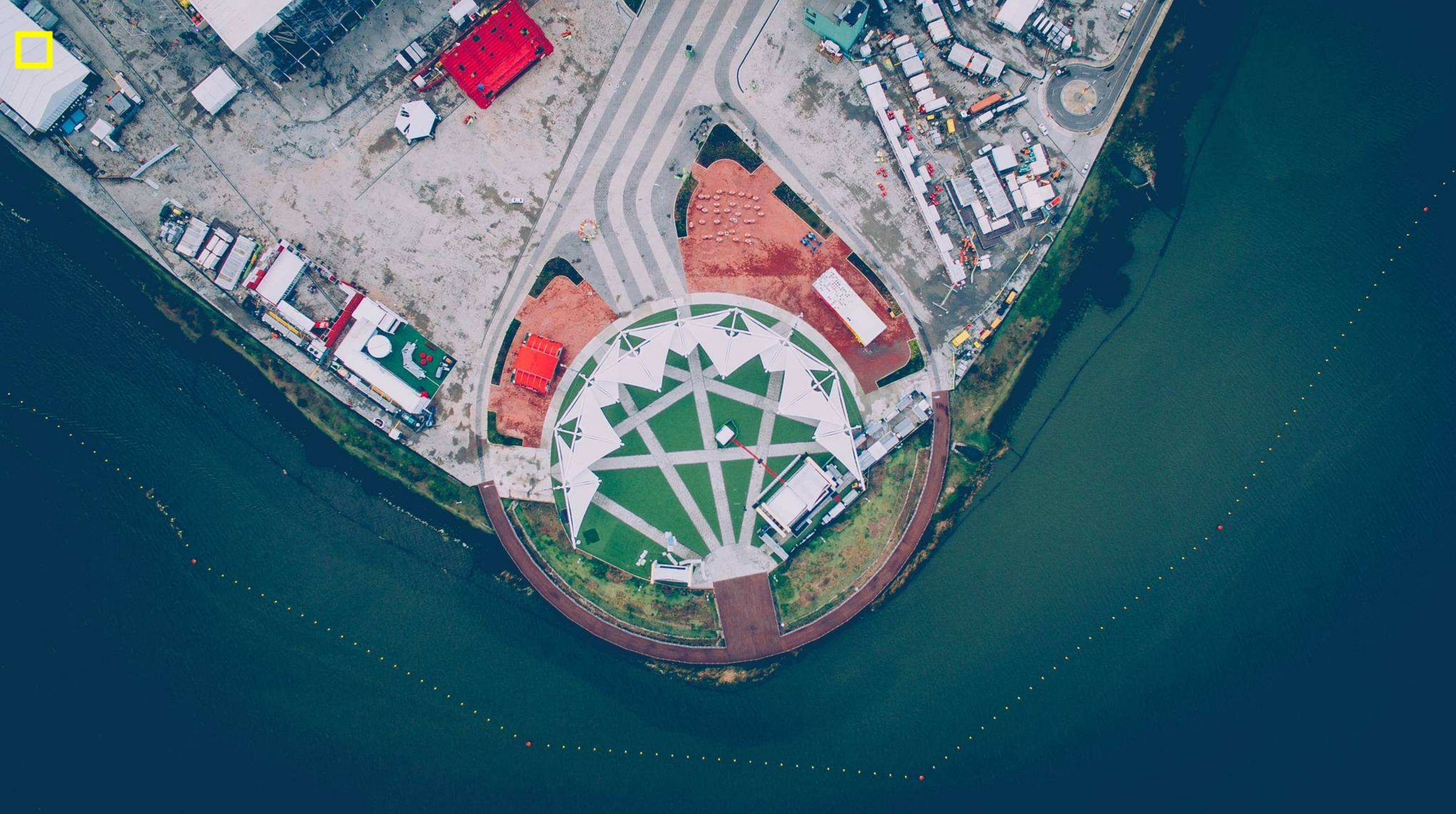}
         \caption{Input image from DIV2K}
         \label{fig:div2k}
     \end{subfigure}
    \begin{subfigure}[b]{0.15\textwidth}
        \centering
         \includegraphics[width=0.8\textwidth]{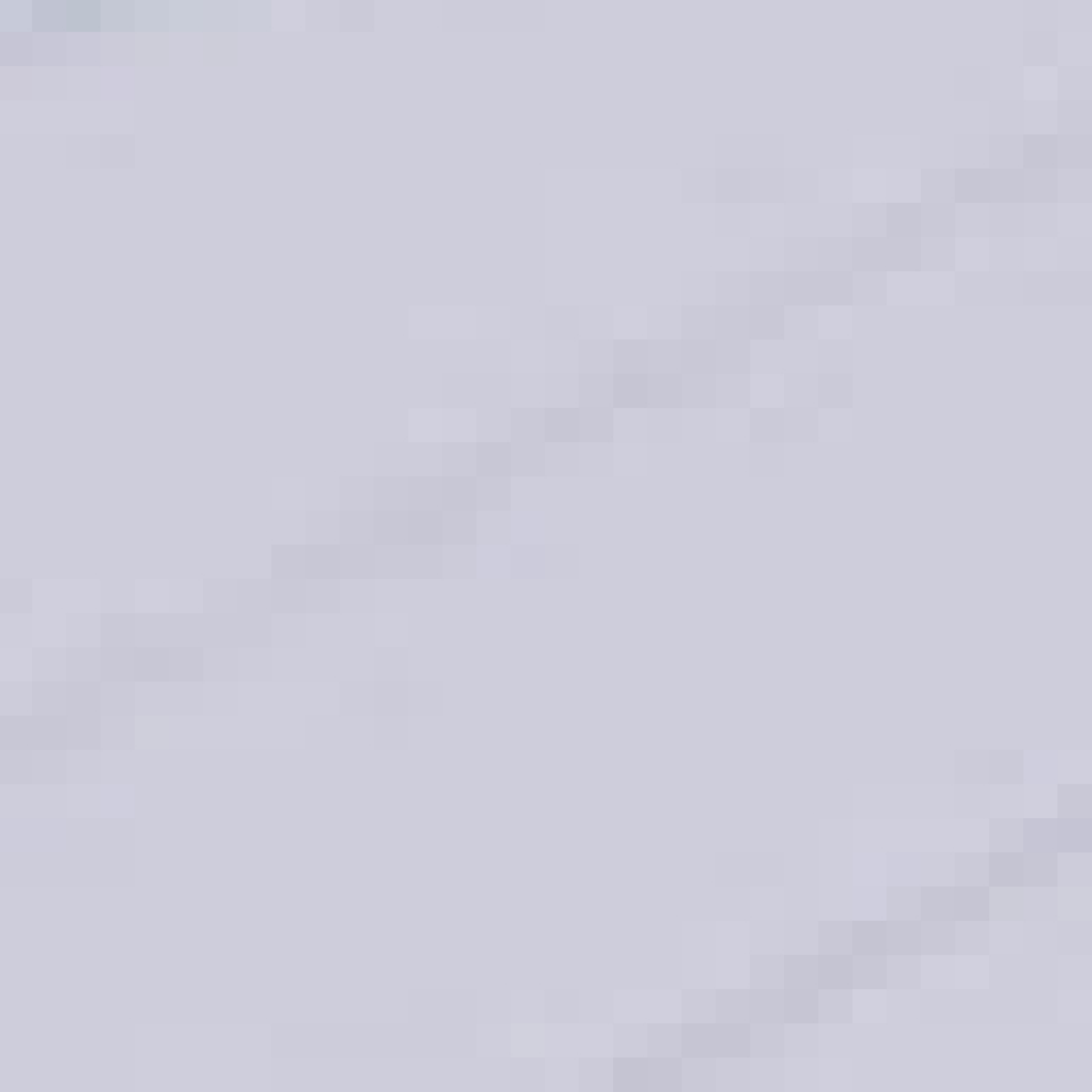}
         \caption{Zoom-in}
         \label{fig:div2kr}
     \end{subfigure}
     \begin{subfigure}[b]{0.15\textwidth}
        \centering
         \includegraphics[width=0.8\textwidth]{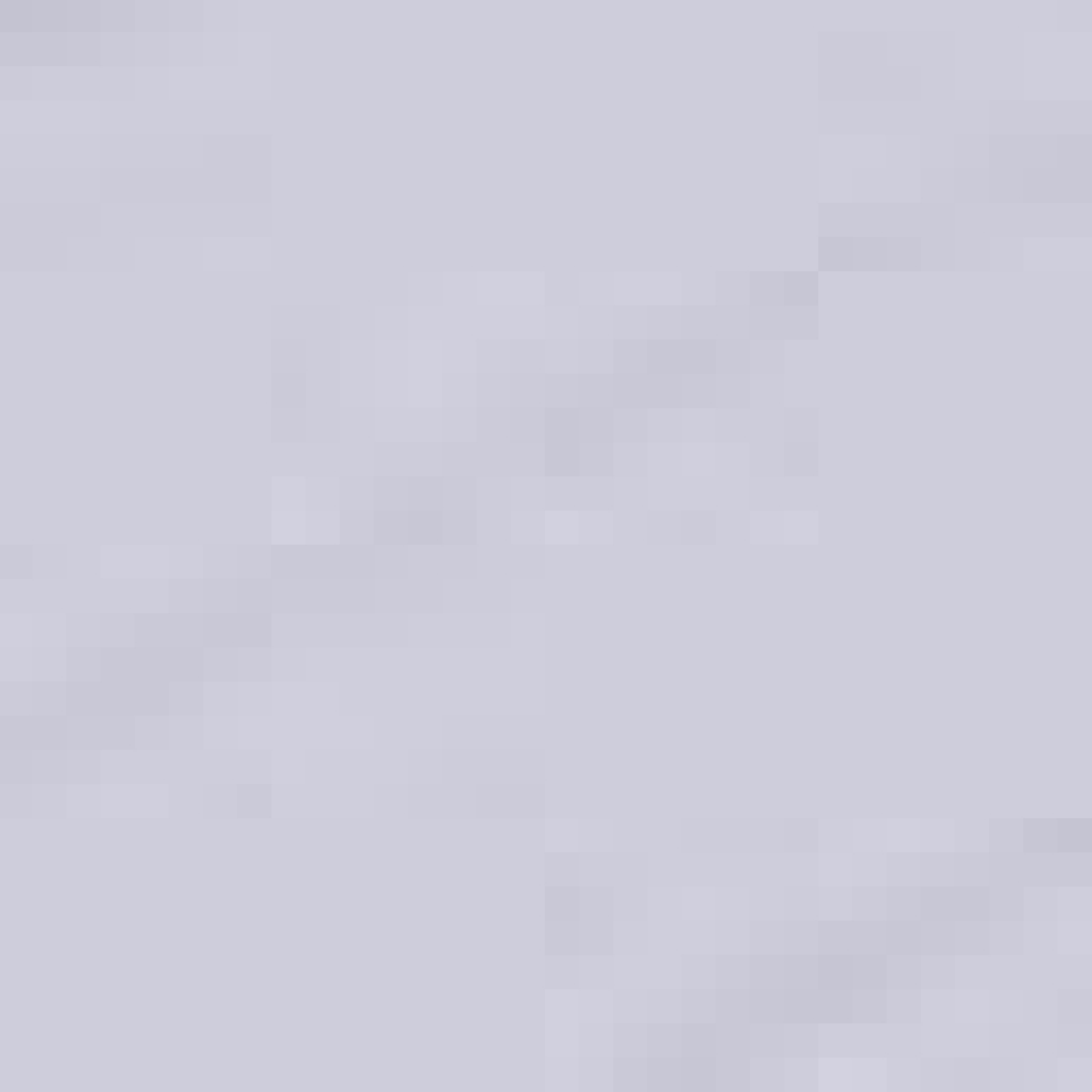}
         \caption{JPEG}
         \label{fig:div2kj}
     \end{subfigure}
     \begin{subfigure}[b]{0.15\textwidth}
        \centering
         \includegraphics[width=0.8\textwidth]{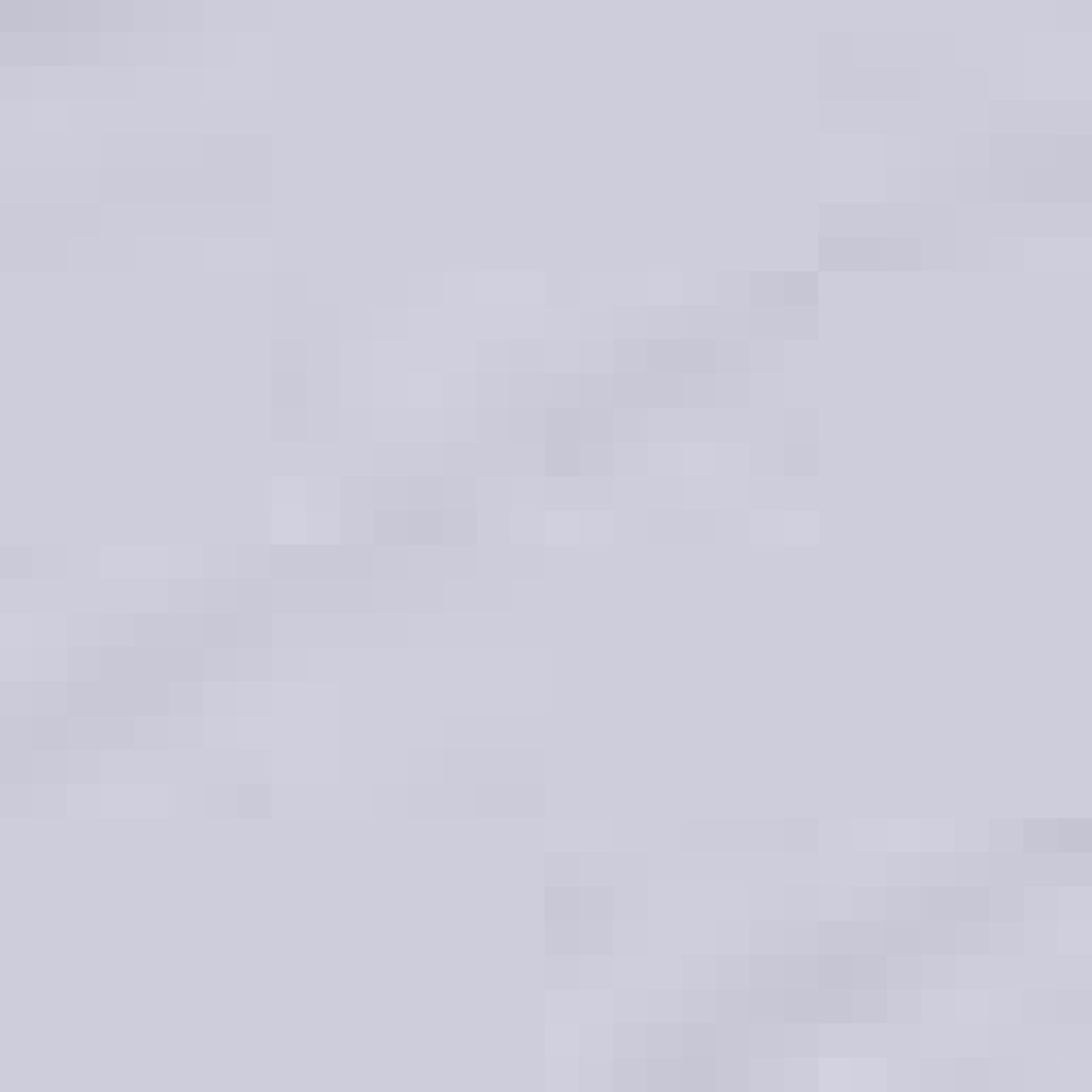}
         \caption{Ours}
         \label{fig:div2ko}
     \end{subfigure}
\caption{Visualization of a zoom-in region from an input DIV2K image, the 
decoded region by JPEG and the proposed method with QF=70.}\label{fig:example_div2k}
\end{figure}
%%%%%%%%%%%%%%%%%%%%%%%%%%%%%%%%%%%%%%%%%%%%%%%%%%

%%%%%%%%%%%%%%%%%%%%%%%%%%%%%%%%%%%%%%%%%%%%%%%%%%
\begin{figure}[htp]
     \centering
    \begin{subfigure}[b]{0.5\textwidth}
        \centering
         \includegraphics[width=0.4\textwidth]{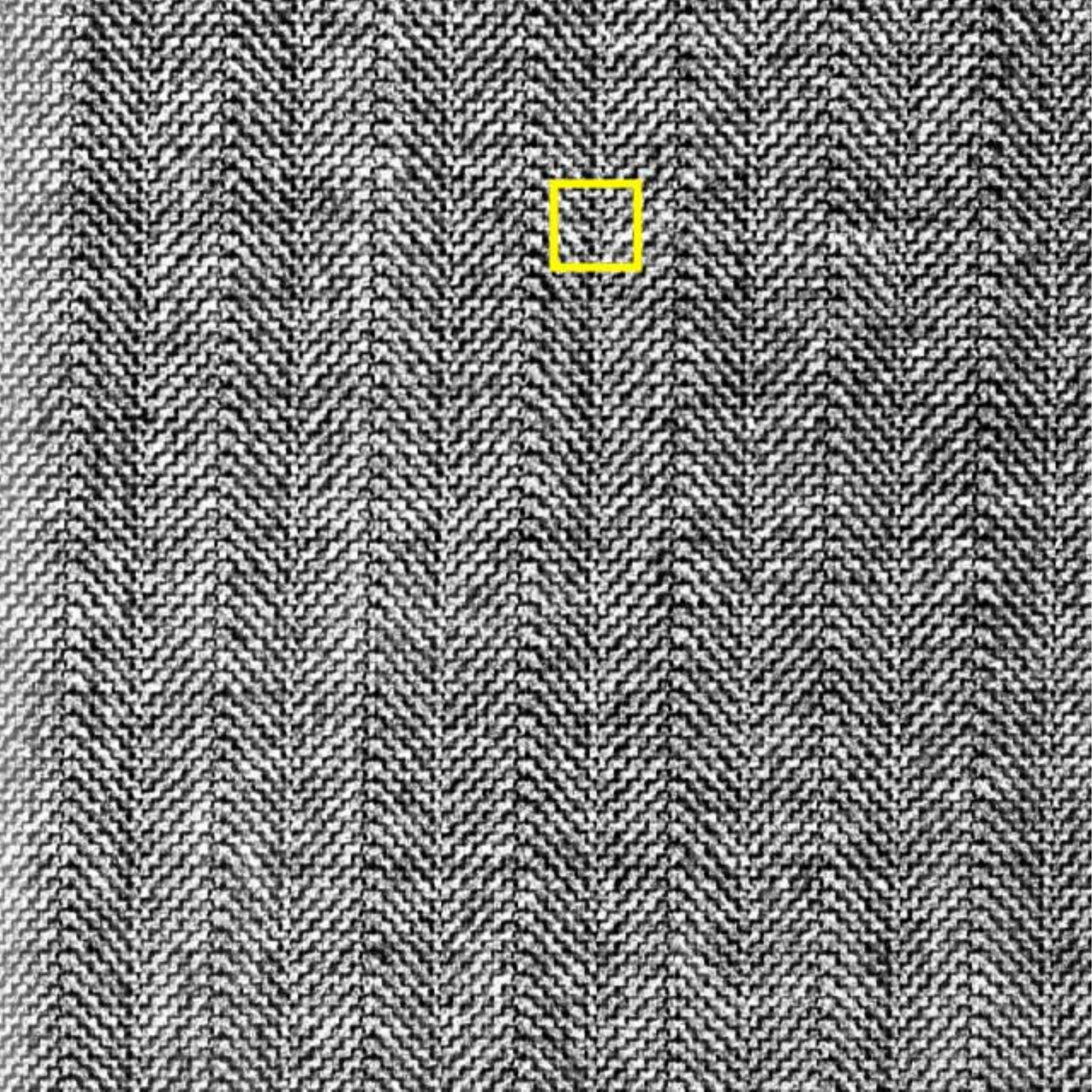}
         \caption{Input image from MBT}
         \label{fig:mbt}
     \end{subfigure}
     \begin{subfigure}[b]{0.15\textwidth}
        \centering
         \includegraphics[width=0.8\textwidth]{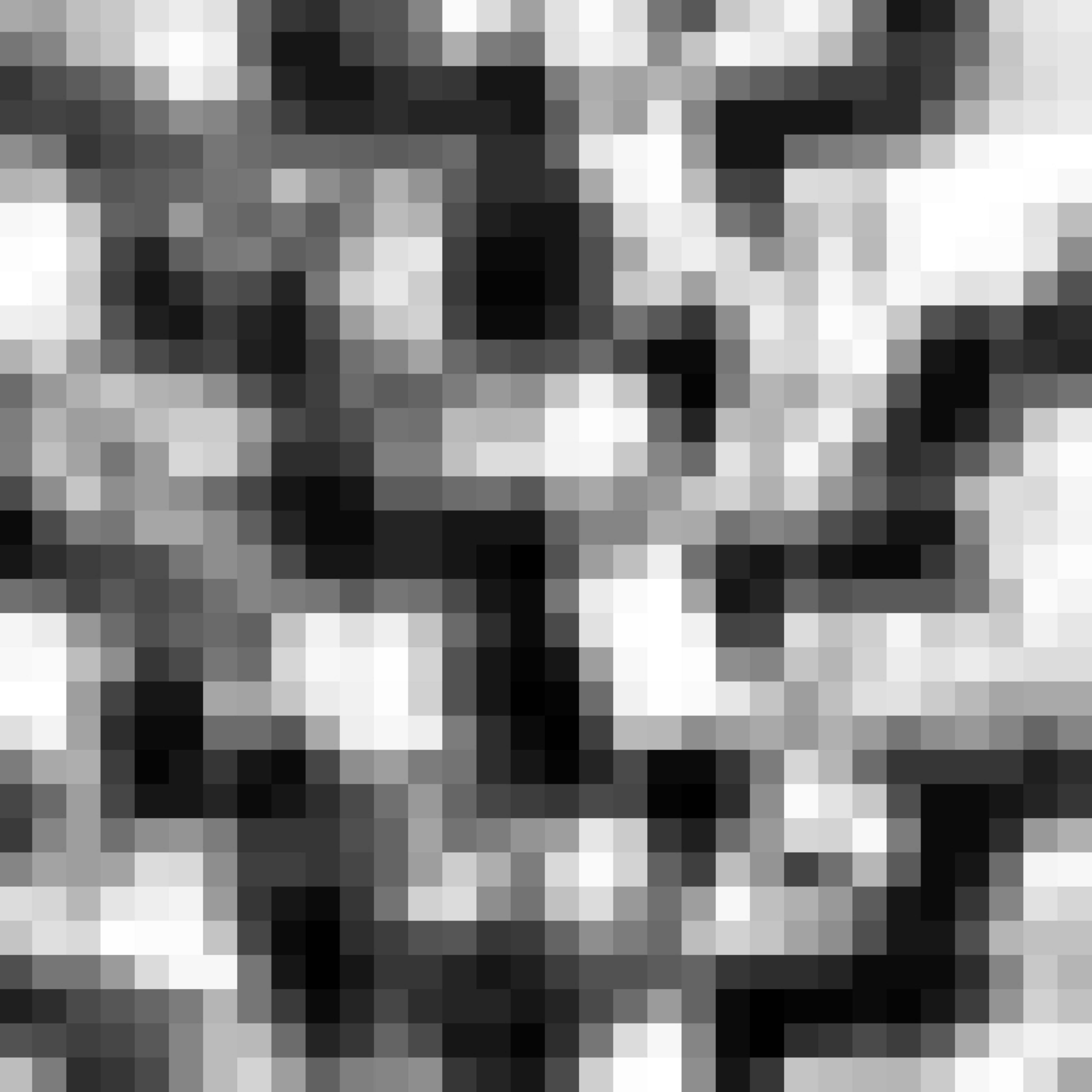}
         \caption{Zoom-in}
         \label{fig:mbtr}
     \end{subfigure}
     \begin{subfigure}[b]{0.15\textwidth}
        \centering
         \includegraphics[width=0.8\textwidth]{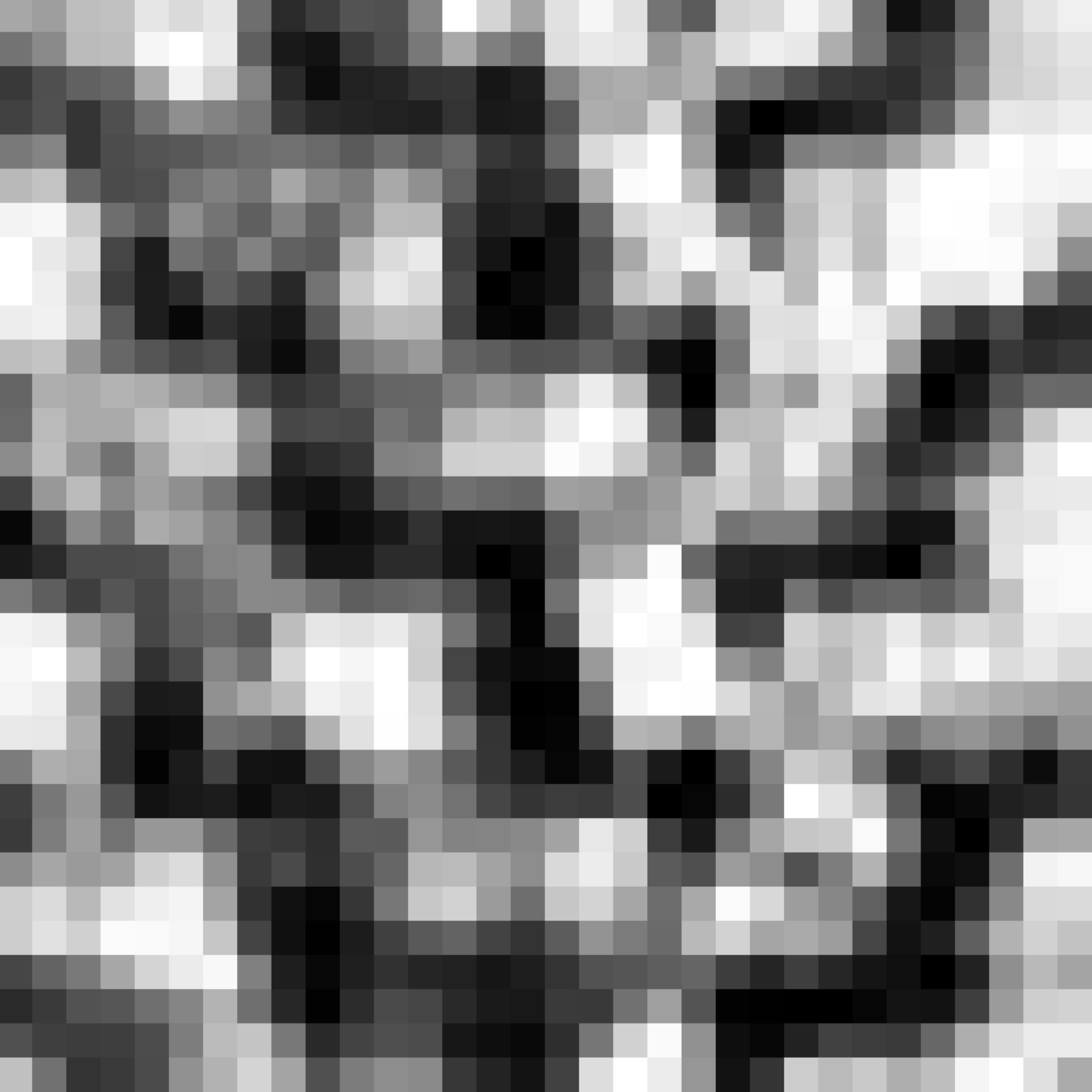}
         \caption{JPEG}
         \label{fig:mbtj}
     \end{subfigure}
     \begin{subfigure}[b]{0.15\textwidth}
        \centering
         \includegraphics[width=0.8\textwidth]{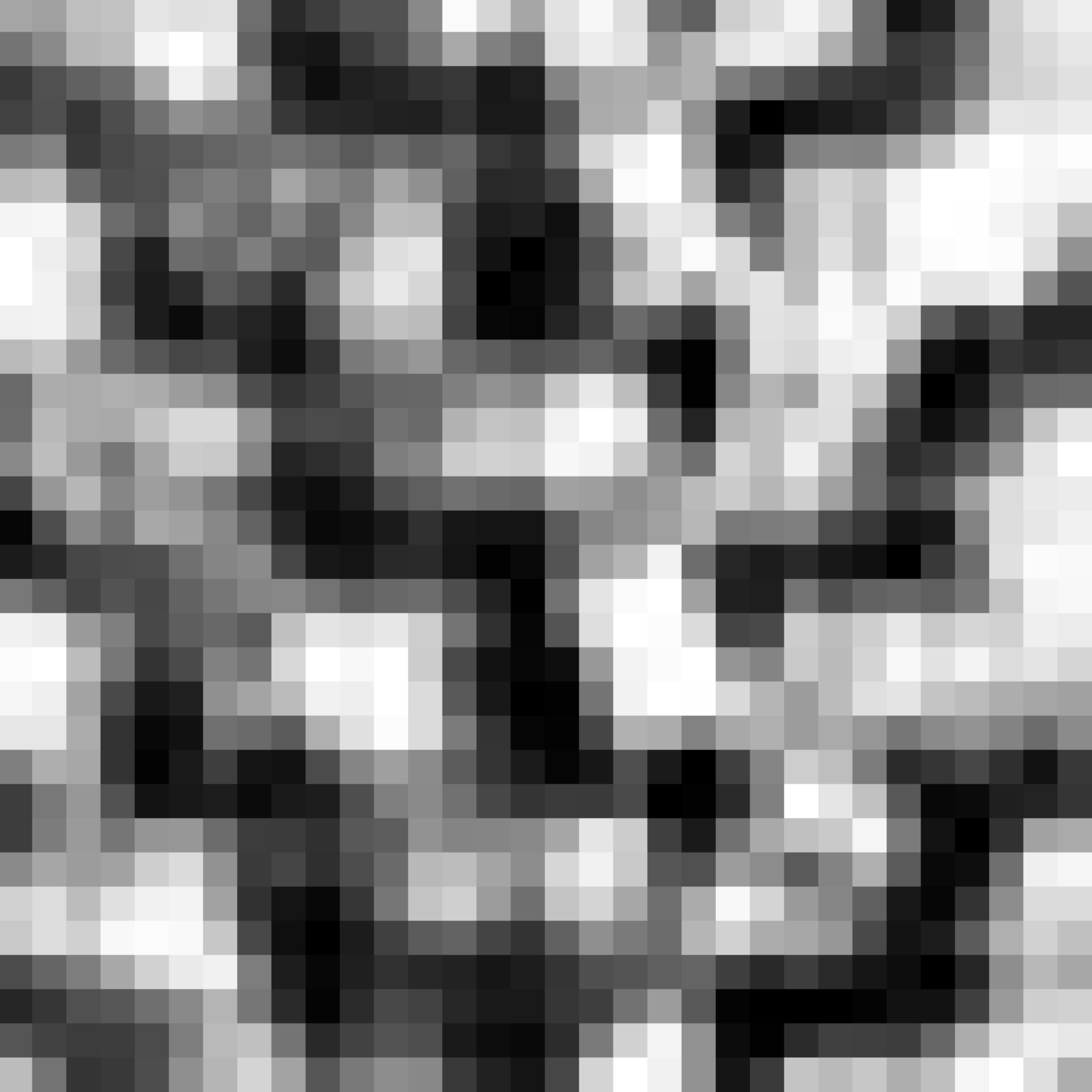}
         \caption{Ours}
         \label{fig:mbto}
     \end{subfigure}
\caption{Visualization of a zoom-in region of an input image from MBT, the
decoded region by JPEG and the proposed method with QF=70.}\label{fig:example_mbt}
\end{figure}
%%%%%%%%%%%%%%%%%%%%%%%%%%%%%%%%%%%%%%%%%%%%%%%%%%

\section{Experiments}\label{sec:experiment}

{\bf Experimental Setup.} We perform experiments on three datasets with
the libjpeg software \cite{libjpeg} on a Unix machine.  The three
datasets are: Kodak \cite{kodak}, DIV2K \cite{div2k}, and Multiband
Texture (MBT) \cite{abdelmounaime2013new}. The first two datasets
contain generic images. They are used to test the power of the proposed
IDCT method in typical images. The third dataset has specific content
and it is desired to see whether our solution can be tailored to it for
more performance gain.  For each dataset, we split images into disjoint
sets - one for training and the other for testing.  There are 24, 785
and 112 total images in Kodak, DIV2K, and MBT datasets, and we choose
10, 100 and 20 from them as training images, respectively. The DCT
remains the same as that in libjpeg. The IDCT kernel is learned at a
certain QF based on discussion in Sec.  \ref{sec:method}. Afterwards,
the learned kernel is used to decode images compressed by libjpeg. Both
RGB-PSNR and SSIM \cite{wang2004ssim} metrics are used to evaluate the
quality of decoded images. 

{\bf Impact of QF Mismatch.} It is impractical to train many kernels
with different QF values. To see the QF mismatch effect on kernel
learning, we show the $L_2$ norm of differences between learned kernels
derived by different QFs in Fig.~\ref{fig:l2}. We see from the figure
that the kernel learned from a certain QF is close to those computed
from its neighbouring QF values except for very small QF values (say,
less than 20). When QF is very small, quantized DCT coefficients have a
lot of zeros especially in high frequency region.  This makes linear
regression poor. The IDCT kernel derived from these quantized
coefficients contains more zeros in each column which makes it different from
others.  We plot the PSNR value and the SSIM value of decoded test
images using the standard IDCT in JPEG and the proposed optimal IDCT as
a function of QF in Fig.  \ref{fig:evaluation}. We see a clear gain
across all QFs and all datasets.  Furthermore, we show the averaged PSNR
and SSIM gains offered by the proposed IDCT designed with the fixed QF
values over the standard DCT in Table \ref{table:comparison}. The PSNR
gain ranges from 0.11-0.30dB. 

%%%%%%%%%%%%%%%%%%%%%%%%%%%%%%%%%%%%%%%%%%%%%%%%%%%%%%%%%%
\begin{table}  
\centering
\begin{tabular}{ccccc}  \hline  
\backslashbox{Dataset}{QF}&	        &50	        & 70  		&90	 \\  \hline  
\multirow{2}{*}{Kodak}	&PSNR (dB)  	&+0.1930 	&+0.2189 	&+0.2057	\\  
		        &SSIM           &+0.0029	&+0.0025 	&+0.0012	\\  \hline 
\multirow{2}{*}{DIV2K}	&PSNR (dB)  	&+0.1229	&+0.1488	&+0.1144	\\  
		        &SSIM           &+0.0024	&+0.0020	&+0.0008	\\  \hline 
\multirow{2}{*}{MBT}	&PSNR (dB)  	&+0.1603 	&+0.2537 	&+0.3038	\\  
		        &SSIM  	        &+0.0025	&+0.0022	&+0.0005	\\  \hline  
\end{tabular}  
\caption{Evaluation results on test images in the Kodak, DIV2K, and MBT
datasets, where training QFs are set to 50, 70 and 90 for each column.}\label{table:comparison}
\end{table} 
%%%%%%%%%%%%%%%%%%%%%%%%%%%%%%%%%%%%%%%%%%%%%%%%%%%%%%%%%%

{\bf Quality Comparison via Visual Inspection.} We show a zoom-in region
of three representative images, each of which is selected from Kodak,
DIV2K and MBT datasets using the standard IDCT and the proposed IDCT for
visual inspection in Figs. \ref{fig:example_kodak},
\ref{fig:example_div2k} and \ref{fig:example_mbt}, respectively. The
proposed IDCT method gives better edge, smooth and texture regions over
the standard IDCT in these three examples. Specifically, edge boundaries
suffer less from the Gibbs phenomenon due to the learned kernel.
Similarly, texture regions are better preserved and the smooth regions
close to edge boundaries are smoother using the learned kernel. All
familiar quantization artifacts decrease by a certain degree. 

%%%%%%%%%%%%%%%%%%%%%%%%%%%%%%%%%%%%%%%%%%%%%%%%%%
\begin{figure}[tp]
 \centering
     \begin{subfigure}[b]{0.33\textwidth}\centering
         \includegraphics[width=0.99\textwidth]{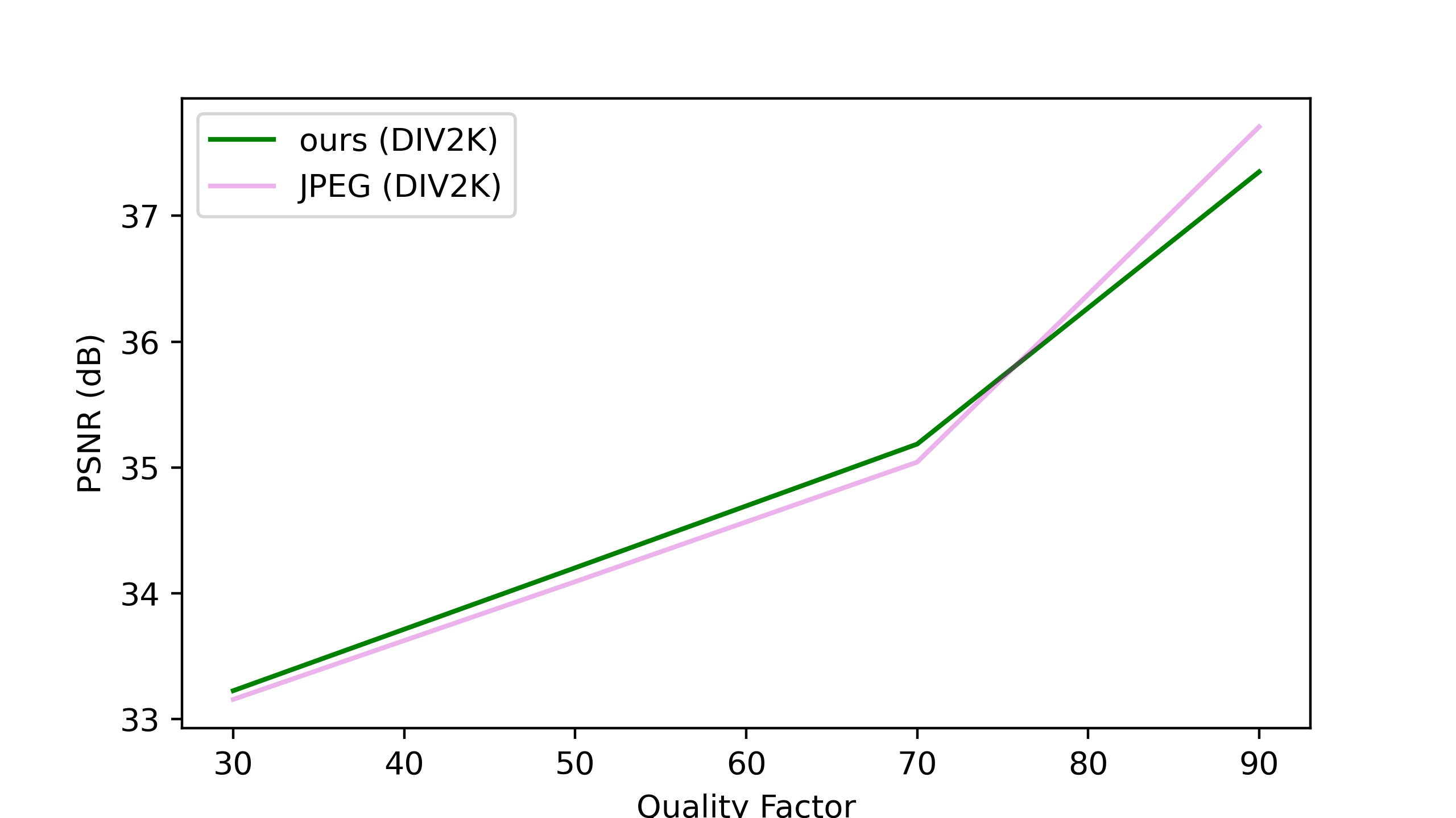}
         \caption{PSNR versus QF}
     \end{subfigure}
     \begin{subfigure}[b]{0.33\textwidth}\centering
         \includegraphics[width=0.99\textwidth]{figs/kodak_ssim.pdf}
         \caption{SSIM versus QF}
     \end{subfigure}
\caption{Comparison of quality of decoded images using the standard
IDCT and the proposed IDCT trained by the Kodak dataset yet tested 
on the DIV2K dataset.}\label{fig:DIV2K_kodak}
\end{figure}
%%%%%%%%%%%%%%%%%%%%%%%%%%%%%%%%%%%%%%%%%%%%%%%%%%

{\bf Impact of Image Content.} Another phenomenon of interest is the
relationship between image content and the performance of the proposed
IDCT method. On one hand, we would like to argue that the learned IDCT
kernel is generally applicable. It is not too sensitive to image
content. To demonstrate this point, we use 10 images from the Kodat
dataset to train the IDCT kernel with QF=70 and then apply it to all
images in the DIV2K dataset. This kernel offers a PSNR gain of 0.24dB
over the standard IDCT kernel. On the other hand, it is still
advantageous if the training and testing image content matches well. As
shown in Table~\ref{table:comparison}, MBT has a higher PSNR gain than
Kodak and DIV2K. It is well known that texture images contain
high-frequency components. When QF is smaller, these components are
quantized to zero and it is difficult to learn a good kernel. Yet, when
QF is larger, high-frequency components are retained and the learned
kernel can compensate quantization errors better for a larger PSNR gain
on the whole dataset. 

\section{Conclusion and Future Work}\label{sec:conclusion}

An IDCT kernel learning method that compensates the quantization effect
was proposed in this work. The proposed method adopts a machine learning
method to estimate the optimal IDCT kernel based on the quantized DCT
coefficients and the desired output block of image pixels. Extensive
experiments were conducted to demonstrate a clear advantage of this new
approach. The learned kernel is not sensitive to the learning QF values
neither to the image content. It offers a robust PSNR gain from 0.1 to
0.3 dB over the standard JPEG. Since it is used in the decoder, it does
increase encoding time or the compressed file size. The learned kernel
can be transmitted offline as an overhead file or simply implemented by
the decoder alone.

It is interesting to consider region adaptive kernels. For example, we
can roughly categorize regions into smooth, edge and textured regions.
The distribution of DCT coefficients in these regions are quite
different.  Thus, we can use a clustering technique to group similar DCT
coefficient distributions and conduct learning within a cluster.
Furthermore, we may consider separable IDCT kernels since they can be
implemented more effectively. Finally, it is desired to apply the same
idea to video coding such as H.264/AVC and HEVC for further performance
improvement. 

\newpage
\bibliographystyle{IEEEtran}
\bibliography{refs}

\end{document}